\documentclass{appolb}
\usepackage{epsfig}
\usepackage{bm}

\begin{document}
\title{Search for new physics in charm processes%
\thanks{Presented at Kazimierz by S. Fajfer}%
}
\author{Svjetlana Fajfer
\address{J. Stefan Institute, Jamova 39, P. O. Box 3000, 1001 Ljubljana, Slovenia and 
Department of Physics,
  University of Ljubljana, Jadranska 19, 1000 Ljubljana, Slovenia}
\and
Ilja Dor\v sner$^a$, Jernej Kamenik$^{a,b}$, Nejc Ko\v snik$^a$
\address{$a$) J. Stefan Institute, Jamova 39, P. O. Box 3000, 1001 Ljubljana, Slovenia\\ $b$) INFN, Laboratori Nazionali di Frascati, Via E. Fermi 40 I-00044 Frascati, Italy
}
}
\maketitle

\begin{abstract}
  Charm procceses are usually not considered to be favorable
  candidates in the search for new physics. Recent disagreement
  between experimental and lattice QCD results on the $D_s$ decay
  constant has motivated us to systematically reinvestigate role of
  leptoquarks in rare charm meson decays.  We include constrains
  coming from the light meson decays.

\end{abstract}
  
\section{Introduction}
Appearance of new physics in the charm processes is known to be
screened either due to long distance contribution, or due to
special interplay of the GIM mechanism and CKM parameters in the charm
processes amplitudes.

The $c \to u$ transition, however, gives a chance to study effects of
new physics in the up-like quark sector. The QCD corrected effective
Lagrangian gives $\rm BR(c\to u\gamma)\simeq3\times10^{-8}$
\cite{WG2}. A variety of models beyond the standard model were
investigated and it was found that the gluino exchange diagrams within
general minimal supersymmetric SM (MSSM) might lead to the enhancement
of the order $10^{2}$. The leading contribution to $c\to ul^+l^-$ in
general MSSM with the conserved R parity comes from one-loop diagram
with gluino and squarks in the loop (e.g. [1]).  It proceeds via
virtual photon and significantly enhances the $c\to ul^+l^-$ spectrum
at small dilepton mass $m_{ll}$.  The experimental results on
parameters describing the $D^0 - \bar D^0$ oscillations have
stimulated studies of new physics in charm sector. Also, it was found
that the result of experimental measurements for the leptonic decay
rates of $D_s$ mesons (see refs.[3-6] in
[2]) 
and the lattice results for the relevant $f_{D_s}$ decay constant
(refs.[7-9] in
[2]) 
disagree by $2.3\,\sigma$, while the corresponding values for $f_D$
are in perfect agreement.

We reinvestigate possible explanation of this puzzle by the leptoquark
mediation.  In our approach leptoquarks transform as a weak
interaction triplet, doublet, or singlet in a model independent
approach. Generally, leptoquarks which also couple to diquarks mediate
fast proton decay and are therefore required to be much above the
electroweak scale, 
making them uninteresting for other low energy phenomena.  ``Genuine''
leptoquarks on the other hand, couple only to pairs of quarks and
leptons, and may thus be inert with respect to proton decay. In such
cases, proton decay bounds would not apply and leptoquarks might
affect low-energy phenomena.  We consider whether light scalar
``genuine'' leptoquarks can explain the $f_{D_s}$ puzzle and at the
same time comply with all other measured flavor observables. Then we
consider a $SU(5)$ GUT model in which the leptoquarks are accommodated
in the $45$-dimensional Higgs representation.  Using the current
experimental measurements in $\tau$, kaon and charm sectors, we find
that scalar leptoquarks cannot naturally explain the $D_s \to \mu \nu$
and $D_s \to \tau \nu$ decay widths simultaneously.

We construct all possible renormalizable scalar leptoquark
interactions with SM matter fields.  There are a few such
dimension-four operators using leptoquarks which are either singlets,
doublets or triplets under the $SU(2)_L$. If we furthermore require
that such leptoquarks contribute to leptonic decays of charged mesons
at tree level, we are left with three possible representation
assignments for the $SU(3)_c\times SU(2)_L\times U(1)_Y$ gauge groups:
$(\mathbf 3, \mathbf 3,-1/3)$, $({\mathbf {\bar 3}}, \mathbf{2},-7/6)$
and $(\mathbf {3},\mathbf {1},-1/3)$. Only the weak doublet leptoquark
is ``genuine'' in the above sense. However, using a $SU(5)$ GUT model
where the relevant leptoquarks are embedded into the 45-dimensional
Higgs representation~($\mathbf {45}_H$), we show how leptoquark
couplings to matter can arise and in particular, how the dangerous
couplings to diquarks -- both direct and
indirect 
-- can be avoided.

In our study we assume the mass degeneracy of the leptoquark multiplets. 
Even more, the  electroweak precision observables do not allow their degeneracy. 
We study only  role of leptoquark mediation 
 at tree level since these already involve
processes forbidden in the SM at tree level, i.e., flavor changing
neutral currents (FCNCs) and lepton flavor violation (LFV)
processes. Finally, since the present $f_{D_s}$ deviation is of mild
significance, we require all the measured constraints to be satisfied
within one standard deviation (at $68\,\%$ C.L.) except upper bounds, for
which we use published $90\,\%$ C.L. limits. 

After the electroweak~(EW) symmetry breaking, quarks and leptons
acquire their masses from their respective Yukawa interactions. Since
these are not diagonal in the weak basis, a physical CKM and PMNS
rotations are present between the upper and the lower components of
the fermion doublets, when these are written in term of the physical
(mass eigen-)states. Consequently, it is impossible to completely
isolate leptoquark mediated charged current interactions to a
particular quark or lepton generation in the left-handed sector {\it
  irrespective of the initial form of the leptoquark couplings to SM
  matter fields, unless there is some special alignment with the
  right-handed quark sector}.
To see this, we denote as $X^{(}{}'{}^{)}$ a $3\times 3$ arbitrary Yukawa
matrix in the weak basis, and write down flavor structure of interaction
of the quark and lepton doublet parts
\begin{equation}
\label{eq:FlavStruct1}
  \overline{Q^w_q} X^{q \ell} = ( \overline{u^w_q} \quad \overline{
    d^w_q})  X^{q \ell}
  = (\overline{u_q} \quad \overline{d'_q}) (U^\dagger X)^{q \ell},
\end{equation}
\begin{equation}
\label{eq:FlavStruct2}
  X'^{q \ell } L^w_\ell = X'^{q \ell} (\nu^w_\ell \quad
    e^w_\ell)^T = (X' E)^{q \ell} (
    \nu'_\ell \quad e_\ell )^T,
\end{equation}
where fields with $w$ superscript are in the weak basis, whereas $d' =
V_{CKM} d$ and $\nu' = V_{PMNS} \nu$. The unitary matrices $U, D, E$,
and $N$ rotate the fields from mass to weak basis and are unphysical
\emph{per se}, so we absorb them in redefinition of the couplings
(e.g. $Y_{LQ} \equiv U^\dagger X$ on the quark and $Y_{LQ}' \equiv X' E$ on the
lepton side) and consider them as free parameters. Explicit feature of
our choice of couplings is that all remaining rotations are assigned
to down-type quark ($V_{CKM} = U^\dagger D$) and neutrino ($V_{PMNS} =
E^\dagger N$) sectors. 
In any case, when one sums the rates for all neutrino
species it becomes evident that in the summed rate, all the neutrino indices
are replaced by the lepton flavors. This is equivalent to the absence
of mixing in the lepton doublets. In what follows, we will use the
convention, where a relative CKM factor is assigned to the down-type
quarks \cite{DFKK}. 

\section{Triplet, doublet and singlet leptoquarks }
The triplet leptoquark interaction
Lagrangian has only one  term
\begin{equation}
  \mathcal L_3 = Y_3^{ij}\, \overline{Q_i^{c}} i \tau_2\, 
 \vec \tau\cdot \vec \Delta_3^* \, L_j + \mathrm{h.c.} \,,
\label{eq:L3}
\end{equation}
where $\overline{Q^c} = -Q^T C^{-1}$, $C = i \gamma^2 \gamma^0$ and
$\tau$ are the Pauli matrices. The $3\times 3$ coupling matrix
$Y_3$ is arbitrary.  In the concrete $SU(5)$ model, the above couplings are due to the contraction of
$\bm{10}$ and $\bm{\overline{5}}$ with $\bm{45}^*_H$ -- also
responsible for giving masses to the down quarks and charged
leptons. We consider constraints coming from the following processes:
$D_s \to \mu (\tau) \nu_{\mu (\tau)}$, $\tau \to \eta,  \phi, \pi, K \nu_\tau$, $K \to \mu \nu_\mu$, $K^+ \to \pi^+ \nu \bar \nu$, $K_L \to \mu^+ \mu^-$ and 
ratios $Br(\tau \to K \nu)/Br(K \to \mu \nu)$  and  
$Br(\tau \to \pi \nu)/Br(\pi \to \mu \nu)$. 
In our calculations \cite{DFKK} enter  the coupling $\tilde Y_3$ which contains  an
additional $V_{CKM}$ rotation for the down-type quarks
\begin{equation}
  \tilde Y_3^{q\ell} \equiv \left\{\begin{array}{ccc}
      Y_3^{q\ell} &;& q = u,c,t,\\
      (V_{CKM}^T Y_3)^{q\ell} &;& q = d,s,b.
    \end{array} \right.
\end{equation}
We use the following parameterization:
\begin{equation}
Y_{LQ}^{q\ell} = y^{\ell}_{LQ} (\sin\phi, \cos\phi).
\end{equation}
\begin{figure}[t]
\begin{center}
\begin{tabular}{c}
\epsfig{file=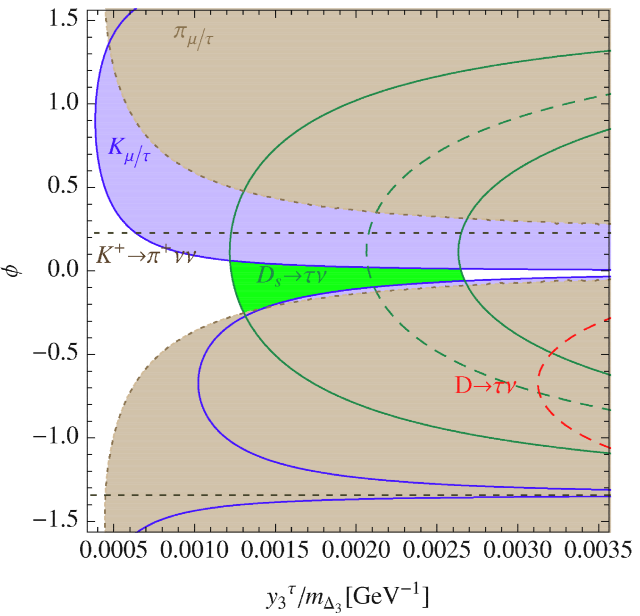,width=7cm}\\
\\
\epsfig{file=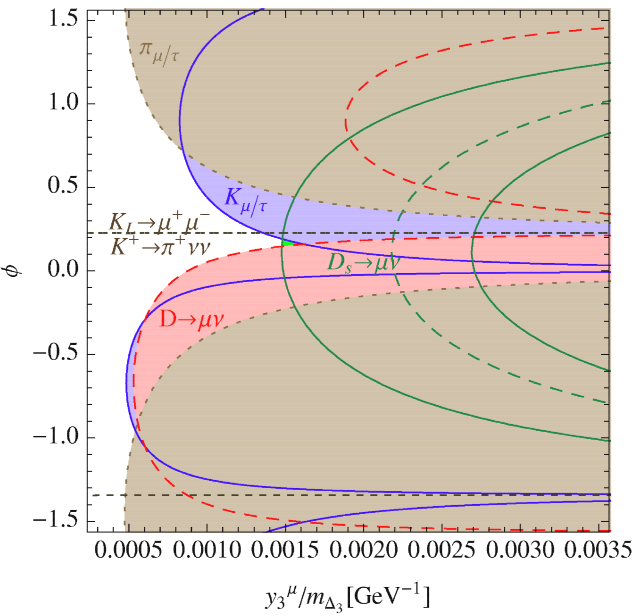,width=7cm}
\end{tabular}
\end{center}
\caption{ Combined bounds on the triplet leptoquark parameters in the
  two-generation limit in the tau (upper plot) and muon (lower plot)
  sectors. All bands represent $68\%$ C.L. exclusion intervals, except
  the upper bound on $D\to\tau\nu$ which is taken at $90\%$ C.L.. The
  $K^+\to\pi^+\nu\bar\nu$ and $K_L\to\mu^+\mu^-$ constraints can only
  be satisfied on the two horizontal dashed lines. Within the bands
  around the dashed line denoted by $D_s \to \ell \nu$, the
  $D_s\to\ell\nu$ excess can be accounted for. \label{fig:1} }
\end{figure}
The constraints presented in  Fig.~\ref{fig:1} 
  clearly disfavors a triplet
leptoquark explanation of the $D_s\to\mu\nu$ excess.

The doublet leptoquarks are innocuous as far as proton decay is
concerned. The allowed dimension four interactions in this case are
\begin{equation}
  \mathcal L_2 = Y_{2L}^{ij}\,  \overline Q_i \,i \tau_2 \Delta_2^*\, e_j  
+ Y_{2R}^{ij}\, {\overline u_i}  \Delta_2^\dagger L_j  + \mathrm{h.c.}\,.
\label{eq:L2}
\end{equation}
In the particular $SU(5)$ model, the term proportional to $Y_{2R}$ stems
from the contraction of $\mathbf{10}$ and $\mathbf{\overline{5}}$ with
$\mathbf{45}^*_H$ while the $Y_{2L}$ term is due to $\mathbf{10}$
and $\mathbf{10}$ being contracted with $\mathbf{45}_H$.
\begin{figure}[t]
\begin{center}
\epsfig{file=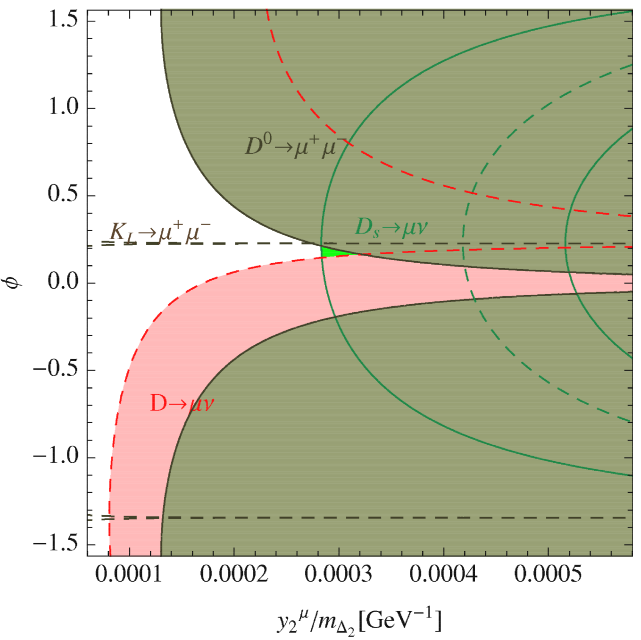,width=7cm}
\end{center}
\caption{ Combined $D^0\to\mu^+\mu^-$ and $D\to\mu\nu$ bounds on the
  doublet leptoquark parameters in the two-generation limit in muon
  sector as explained in the text. $D\to\mu\nu$ band represent $68\%$
  C.L. exclusion interval, while the upper bound on $D^0\to\mu^+\mu^-$
  is taken at $90\%$ C.L..  The dashed line represents central value
  of the $D_s\to\ell\nu$ excess, while neighbouring solid lines
  represent its one-sigma error. However, the bound coming from $K_L \to
  \mu^+ \mu^-$ lies out of this region, making the doublet of
  leptoquark inadequate for the explanation of the $D_s$
  puzzle. \label{fig:2} }
\end{figure}
In our study we use bounds coming from the following decay rates: $D_s
\to \ell \nu $, $D^0 \to \mu^+ \mu^-$ and $K_L \to \mu^+
\mu^-$. Recently, Belle collaboration has presented new bound on the
decay rate $Br(D^0 \to \mu^+ \mu^-) < 1.4 \times 10^{-7}$.  Including
this bound we find that the doublet leptoquark cannot explain the
$D_s$ puzzle. As one sees from the Fig. 2 the bound coming from the
$K_L \to \mu^+ \mu^-$ does not pass the region allowed by other
constraints.

The relevant most general Lagrangian describing singlet 
leptoquarks has two terms:
\begin{equation}
  \mathcal L_1 = Y_{1L}^{ij} \overline{Q_i^c} i \tau_2 \Delta_1^* L_j  + Y_{1R}^{ij} \overline{u_i^c}  \Delta_1^* e_j + \mathrm{h.c.}\,.
  \label{eq:L1}
\end{equation} 
In order to constrain leptoquark couplings we use: $Br(K^+ \to \pi^+
\nu \bar \nu)$, the Belle collaboration bound for $D^0 \to \mu^+
\mu^-$, and the ratios of experimental rates and the standard model
values for the decay $D_s \to \tau \nu_\tau$ and $\tau \to K
\nu_\tau$.  The lack of experimental information on up-quark FCNCs
involving only tau leptons leaves the verdict on the singlet
leptoquark contribution to the $D_s\to\tau\nu$ decay width open. What
is certain is that due to the $K^+\to\pi^+\nu\bar\nu$ constraint any
such contribution has to be aligned with the down-type quark Yukawas
such that $\tilde Y_1^{d\tau}\approx 0$ can be ensured.

By performing a numerical
fit of ($y_1^{\mu},\omega,\phi$) to the above mentioned  constraints we obtain the
result, that the experimental value for $Br(D_s\to\mu\nu)$ cannot be
reproduced within one standard deviation without violating any of the
other constraints, thus excluding the singlet leptoquark as a natural
explanation of the $D_s\to\mu\nu$ puzzle. Same conclusions can be
drawn for the $R$-parity violating minimal supersymmetric SM, where
the interaction term of a down squark to quark and lepton doublets is
present and corresponds to first term in (\ref{eq:L1}), while the
second term is absent in that case.

Existing constraint coming from precision kaon, tau, and $D$ meson
observables imply in a model independent way, that a single scalar
leptoquark cannot explain enhanced $D_s\to\ell\nu$ decay widths for
both $\ell = \mu$ and $\tau$.

\end{document}